\documentstyle[preprint,prb,aps,epsfig]{revtex}

\begin{document}

\title{ Effective Interaction Hamiltonian of Polaron Pairs in
Diluted Magnetic Semiconductors}

\author{D. E. Angelescu and R. N. Bhatt}

\address{Department of Physics and Electrical Engineering, Princeton University, Princeton, NJ 08544} 

\date{\today}


\maketitle
\begin{abstract}
The magnetic interaction of a pair of bound magnetic polarons
(BMP) in diluted magnetic semiconductors (DMS) is analyzed
\emph{via} a generalized Hubbard-type Hamiltonian for two
carriers in the presence of effective magnetic fields arising from
the magnetic polarization of their respective
polarons. For the case where the magnetic fields at the two
sites have equal magnitude but are allowed to have arbitrary
directions, it is shown that the energy of the two polarons
is minimized for a \emph{ferromagnetic} configuration
of the carrier spins (in contrast to the case of hydrogenic
centers in nonmagnetic semiconductors)
if polaron fields are strong enough.
A modified Heisenberg-type Hamiltonian is constructed
to describe the low energy states of the resulting system.   
\end{abstract}

\section{Introduction}
Shallow impurities in doped semiconductors \cite{kohn:ssp}
can be described in terms of a hydrogenic Hamiltonian with
an effective mass given by the band mass, and a Coulomb
potential screened by the dielectric constant of the host
semiconductor. While the``effective
mass'' equation is slightly more complicated
for donors in indirect band-gap semiconductors,
and for acceptors a matrix
version of a generalized Hydrogen problem is obtained, these
differences change details, but not the
basic physics. Therefore, the hydrogenic model is a
useful guide for studying and understanding interactions
between impurities in doped semiconductors
 \cite{bhatt:physcr}$^,$\cite{bhatt:phrevlet} .   
At low concentrations the interactions between impurity
centers can be modelled in terms of an exchange Hamiltonian
involving, as the dominant term \cite{mattis:magn}, pairwise
Heisenberg exchange corresponding to the Hydrogen molecule problem
\cite{ashcroft:ssp}, where the exchange interactions
are known to be anti-ferromagnetic at all distances \cite{Herring:}.

In diluted magnetic semiconductors (DMS), a small fraction 
of the nonmagnetic ions
that form the lattice are replaced by magnetic ions
such as Mn or Fe. Several features (such as variable band
gaps, optical response, spin polarized transport, as well as the
unusual magnetic behaviour analyzed in this paper) have
turned DMS into a topic of considerable interest during
recent years. In the low doping regime (\emph{i.e.}
carrier density below the Mott transition), the susceptibility
(\emph{i.e.}, $dM/dH$)
\emph{vs.} magnetic field ($H$) curve
of such a DMS displays a
curious double-step feature \cite{liu:steps}. To understand
the properties of DMS with dopants, it is not only necessary
to understand the direct interactions between the dopants,
but also the interactions with the magnetic ions which by
themselves contain low lying degrees of freedom. When the
magnetic ions are dilute, their direct interactions are
unimportant. Thus, for example, the problem of a single
shallow impurity in a DMS is well described in terms of an
exchange interaction between the bound carrier (electron or
hole) and the magnetic ion, and is known from extensive
studies to lead to the formation of a bound magnetic polaron (BMP)
\cite{wolff:review}. The spins of the magnetic impurity ions
within one effective Bohr radius of a dopant interact
\emph{via} a sizable exchange with the carrier, thus becoming
aligned and forming large-spin polarons. The polarons align
with an external magnetic field before the individual
magnetic ions do, thus giving rise to the two-step
susceptibility curve. By analyzing the step due to BMPs
(fitting it to a Curie-Weiss form), a ferromagnetic
interraction between the polarons can be deduced. However
this result seems puzzling: in non-magnetic semiconductors
carrier virtual hopping invariably yields anti-ferromagnetism
\cite{bhatt:physcr} .

The problem was analyzed by Durst, Bhatt and Wolff
\cite{bhatt:polarons}$^,$\cite{bhatt1:polarons}. In their work
they showed that a ferromagnetic interraction between the
polarons can be obtained if one considers the overlap of two
polarons formed around two dopants. The two carriers compete
over the spins in the overlapping region. For certain values
of the model parameters the indirect carrier-magnetic
ion-carrier interaction becomes stronger than the direct
carrier exchange, and the polarons align resulting in
ferromagnetism. 

The current work approaches the problem from a different
perspective. The polarons are considered as non-overlapping,
and their effect on the dopant atom is taken into account
through a local magnetic field $h$. A system of two such
polarons is analyzed \emph{via} a generalized Hubbard-type
Hamiltonian, where hopping (matrix element $t$) and Coulomb
interaction (energy U) are turned on. Several cases are
considered (dopants with a single bound state and with
several bound excited states). Ferromagnetic behaviour is
seen to emerge when the carrier is allowed to hop between
the ground state of one dopant atom and excited states of the
other dopant. Numerical work supports the conclusion that
such a ferromagnetic interaction is indeed possible in
realistic conditions. We then discuss the applicability of
a  Heisenberg-type model for two interacting
polarons. 
In the moderately high field domain $t \ll h \ll U$,
an effective Heisenberg Hamiltonian is found
which contains a mixing of the magnetic fields at
the two sites.  

\section{The model and the approach}
Our model consists of two identical atoms, with several bound
states, in arbitrary local magnetic fields. We allow for
hopping of the carriers between the two atoms. The local
magnetic fields represent the exchange fields due to
the magnetic ions at each site. As the number of magnetic
ions around each site is large, and the doping is considered uniform, 
we assume that the magnitudes of 
the magnetic fields at the two sites are equal.
However, the directions of the two fields are
allowed to be arbitrary. Thus, the Hamiltonian we study
has the general form:

\begin{equation}\begin{array}{lc} H = -\frac 12(\nabla _1^2+\nabla _2^2)-h{\bf \hat{h}}(r_1){\bf \cdot S}_1-h{\bf \hat{h}}(r_2)\cdot {\bf S}_2-\frac 1{r_{a1}}-\frac 1{r_{a2}}-\frac
1{r_{b1}}-\frac 1{r_{b2}}+\frac 1{r_{12}}
\end{array}
\label{allH}
\end{equation}
where $h$ is the magnitude of the field, ${\bf \hat{h}(r)}$ is the (arbitrary) direction, $a$ and $b$ are the labels for the two Hydrogenic centers and $1$ and $2$ are the labels for the two electrons.

If we consider a Hubbard-like approximation\cite{ashcroft:ssp} with one energy level per impurity site and no magnetic fields, the Hamiltonian becomes (in second quantized form) : 
\begin{equation}
H=\sum_{\alpha=a,b}(\epsilon({\bf n}_{\alpha\uparrow}+{\bf n}_{\alpha\downarrow})+U{\bf n}_{\alpha\uparrow}{\bf n}_{\alpha\downarrow})+\sum_{{\bf s}=\uparrow,\downarrow}t({\bf c}_{a{\bf s}}^\dagger {\bf c}_{b{\bf s}}+{\bf c}_{b{\bf s}}^\dagger {\bf c}_{a{\bf s}})
\label{Hubbnoh}
\end{equation}
where $a,b$ are labels for the two impurity sites, ${\bf c}_{a{\bf \uparrow}}$ is the annihilation operator for the state on impurity $a$ with up-spin, ${\bf n}_{a{\bf \uparrow}}$ is the occupation number, ${\bf n}_{a{\bf \uparrow}}={\bf c}^\dagger_{a{\bf \uparrow}}{\bf c}_{a{\bf \uparrow}}$ \emph{etc.}

If we also introduce arbitrary number of energy levels on each impurity atom, the Hubbard Hamiltonian (\ref{Hubbnoh}) turns into : 
\begin{equation}
\begin{array}{lc}
\label{fullHubb1}
H=\sum _{i,\alpha,{\bf s}} \epsilon_{i\alpha} {\bf n}_{i\alpha{\bf s}} 
+\sum_{i<j,{\bf s}_1,{\bf s}_2,\alpha}U_{ij}  {\bf n}_{i\alpha {\bf s}_1}{\bf n}_{j\alpha {\bf s}_2}+\\ 
\\
+\sum_{i,\alpha}U_{ii}  {\bf n}_{i\alpha \uparrow}{\bf n}_{i\alpha \downarrow}+\sum_{i,j,\mathbf{s}}t_{i,j}({\bf c}_{jb {\bf s}}^\dagger {\bf c}_{ia {\bf s}}+{\bf c}_{ia {\bf s}}^\dagger {\bf c}_{jb {\bf s}} )
\end{array}
\end{equation}
where $\alpha\in \{a,b\}$ indexes the impurity sites; $i,j$ the atomic levels on each impurity; ${\bf s}_1,{\bf s}_2\in \{\uparrow,\downarrow\}$ the spin degree of freedom; $\epsilon_i$ the energy of level $i$; and $U_{ij}$ the Coulomb interaction energy of electrons in states $i$ and $j$ on the same impurity atom.

Finally, if at each site we consider the arbitrary magnetic fields ${\bf h}_{a}, {\bf h}_b$, and we quantize spin along the axes of the local magnetic fields (\emph{i.e.} ${\bf c}_{ia\uparrow}^\dagger$ creates an electron in the $i$-th state on impurity $a$ with spin parallel to $\mathbf{h}_a$), the Hubbard Hamiltonian becomes :
\begin{equation}
\begin{array}{lc}
\label{fullHubb}
H=\sum _{i,\alpha,{\bf s}} \epsilon_{i\alpha} {\bf n}_{i\alpha{\bf s}} 
+\sum_{i<j,{\bf s}_1,{\bf s}_2,\alpha}U_{ij}  {\bf n}_{i\alpha {\bf s}_1}{\bf n}_{j\alpha {\bf s}_2}
+\sum_{i,\alpha}U_{ii}  {\bf n}_{i\alpha \uparrow}{\bf n}_{i\alpha \downarrow}+\\ 
\\+\sum_{i,\alpha}h_{\alpha}({\bf n}_{i\alpha\uparrow}-{\bf n}_{i\alpha\downarrow})+\sum_{i,j,\mathbf{s}_1,\mathbf{s}_2}t_{i{\bf s}_1,j\mathbf{s}_2}({\bf c}_{jb \mathbf{s}_1}^\dagger {\bf c}_{ia \mathbf{s}_2}+{\bf c}_{ia \mathbf{s}_2}^\dagger {\bf c}_{jb \mathbf{s}_1} )
\end{array}
\end{equation}
We caution the reader that in this case the transition matrix elements $t_{i{\bf s}_1,j\mathbf{s}_2}$ become dependent on the angle $\Theta$ between the two magnetic fields ${\bf h}_a, {\bf h}_b$. We will discuss the relationship between Eqs. \ref{fullHubb},\ref{fullHubb1},\ref{Hubbnoh} and Eq. \ref{allH} in more detail in the concluding section. 

Several models of
increasing complexity were considered : atoms with a single
bound state and without magnetic fields (sec. \ref{reghub}),
atoms with a single bound state in arbitrary magnetic fields
(sec. \ref{reghubfields}), atoms with several excited states
in arbitrary magnetic fields (sec. \ref{2levhub}). The ground
state of the two-center system is shown to undergo a
transition from an antiferromagnetic state (singlet) to a fully
ferromagnetic (triplet) configuration with the increase of the
effective polaron magnetic field (sec. \ref{gsmagn}). The
results for a regular Heisenberg Hamiltonian where the two
spins are in arbitrary fixed fields are calculated as well
and compared with those derived from our
model (sec. \ref{regheis}). Finally we find a modified
Heisenberg type Hamiltonian that agrees with our model in the
moderately high field regime (sec. \ref{modif}).   

\section{Regular Hubbard Model}

\subsection{Regular Hubbard model in zero field}
\label{reghub}
The Hubbard model of the hydrogen molecule \cite{ashcroft:ssp} (see Eq. \ref{Hubbnoh})
consists of two hydrogenic (one-electron) centers, each
with one single electron bound state
of energy $\epsilon$. Electrons are allowed
to hop between the two sites, subject to the restrictions 
imposed by the Pauli principle with a hopping matrix element $t$.
Each center also has one two-electron state, with
energy $2 \epsilon + U $, where $U$ represents the
interaction energy between two electrons on the same atom.

In this case we can decouple the spin-$1$ and spin-$0$
subspaces (since there is no connection between states of
different spin). It turns out that the spectrum of
eigenvalues is $2\epsilon$ (triple degenerate) for spin-$1$
and $2\epsilon-\frac{4t^{2}}{U}$, $2\epsilon+U$ and
$2\epsilon+U+\frac{4t^{2}}{U}$ for spin-$0$. The ground state
therefore has spin $0$ ({\em i.e.} the interaction between the
electrons can be thought of as anti-ferromagnetic). 

\subsection{Regular Hubbard model with arbitrary fields}
\label{reghubfields}
If two arbitrary fixed fields (of strengths $h_a$ and $h_b$
and making an angle $\Theta$) are applied at the two sites,
the analysis becomes more complicated (see Eq. \ref{fullHubb}, and consider that there is a single level, $i=1$, on each impurity). We can quantize the
spins along the axes of the fields, and we can pick as a
basis: \[\{a\uparrow a\downarrow,a\uparrow
b\uparrow,a\uparrow b\downarrow,a\downarrow
b\uparrow,a\downarrow b\downarrow, b\uparrow b\downarrow\}\] 
Each element in the basis is anti-symmetrized, for example
\[a\uparrow b\downarrow \equiv
\frac{1}{\sqrt{2}}(|\psi_a(x_1)\uparrow\rangle|\psi_b(x_2)\downarrow\rangle-|\psi_b(x_1)\downarrow\rangle|\psi_a(x_2)\uparrow\rangle)\]
This basis obviously is not formed of eigenstates of the
total spin, but only of the $z$-component of the spin. 

The new Hamiltonian matrix (ignoring $\mu_B$, the Bohr
magneton, for simplicity) is shown in Eq. \ref{HHubb1} below :

\begin{equation}
\label{HHubb1}
H_1=\left( 
\begin{array}{lccccr}
2\epsilon+U&it\sin\frac{\Theta}{2}&t\cos\frac{\Theta}{2}&-t\cos\frac{\Theta}{2}&-it\sin\frac{\Theta}{2}&0\\

-it\sin\frac{\Theta}{2}&2\epsilon+\frac{h_a+h_b}{2}&0&0&0&-it\sin\frac{\Theta}{2}\\

t\cos\frac{\Theta}{2}&0&2\epsilon+\frac{h_a-h_b}{2}&0&0&t\cos\frac{\Theta}{2}\\

-t\cos\frac{\Theta}{2}&0&0&2\epsilon-\frac{h_a-h_b}{2}&0&-t\cos\frac{\Theta}{2}\\

it\sin\frac{\Theta}{2}&0&0&0&2\epsilon-\frac{h_a+h_b}{2}&it\sin\frac{\Theta}{2}\\

0&

it\sin\frac{\Theta}{2}&t\cos\frac{\Theta}{2}&-t\cos\frac{\Theta}{2}&-it\sin\frac{\Theta}{2}&2\epsilon+U
\end{array}
\right)
\end{equation}

We next make the simplifying assumption that $h_a=h_b=h$.
For the case of DMS, since the magnetic ion distribution
is random, this can be justified if
each polaron has several (N) magnetic
ions producing the exchange field on the carrier, so
$ \vert h_a - h_b \vert = h/\sqrt{N}  \ll h $. 
In this case we obtain two pairs of degenerate
states $\{a\uparrow a\downarrow,b\uparrow b\downarrow\}$ and
$\{a\uparrow b\downarrow,a\downarrow b\uparrow\}$. By making
a $45$ degree rotation within each of the degenerate
subspaces, and by multiplying some of the basis vectors by
$i$ when necessary, we single out two of the eigenvalues
($2\epsilon$ and $2\epsilon+U$), the rest of the matrix
having the simpler form in Eq. \ref{HHubbard1} \mbox{below :}

\begin{equation}
\label{HHubbard1}
H_1\rightarrow\left( 
\begin{array}{lccr}
2\epsilon+h&0&0&\sqrt{2}t\sin\frac{\Theta}{2}\\
0&2\epsilon&0&-2t\cos\frac{\Theta}{2}\\
0&0&2\epsilon-h&\sqrt{2}t\sin\frac{\Theta}{2}\\

\sqrt{2}t\sin\frac{\Theta}{2}&-2t\cos\frac{\Theta}{2}&\sqrt{2}t\sin\frac{\Theta}{2}&2\epsilon+U
\end{array}
\right)
\end{equation}

This Hamiltonian can be solved by perturbation theory (PT).
In the limit $h\rightarrow 0$ the upper three states become
degenerate and the problem needs to be handled by degenerate
perturbation theory. We will not investigate this limit any
further. In the high field limit however, the magnetic field
removes the degeneracy and we can obtain the eigenvalues to
second order by regular PT. Thus we obtain for the lowest
eigenvalues : 

\begin{equation}
\label{EHubbard1}
\left(
\begin{array}{llll}
2\epsilon+h-\frac{2t^{2}\sin^{2}\frac{\Theta}{2}}{U-h}&
2\epsilon&
2\epsilon -\frac{4t^{2}\cos^{2}\frac{\Theta}{2}}{U}&
2\epsilon-h-\frac{2t^{2}\sin^{2}\frac{\Theta}{2}}{U+h}
\end{array}
\right)
\end{equation}

\subsection{Heisenberg Hamiltonian with arbitrary fields}
\label{regheis}
By solving the same problem (two atoms in fixed external
fields) using a Heisenberg Hamiltonian, and comparing the
eignevalues with the results obtained above in Eq.
\ref{EHubbard1}, one can see how the effective exchange
parameter in the Heisenberg Hamiltonian is affected by
external magnetic fields. We start with \cite{explanation:} : 
\begin{equation}
\label{heis}
H_{H1}=\mathbf{h_a\cdot s_a}+\mathbf{h_b\cdot
s_b}+J\mathbf{s_a\cdot s_b}
\end{equation}
We can again quantize the spins along the axes of the two
fields, and work in the basis

\[\{a\uparrow b\uparrow,a\uparrow b\downarrow,a\downarrow b\uparrow,a\downarrow b\downarrow\}\]
which yields the Hamiltonian matrix :
\begin{equation}
\label{HHeisenberg}
H_{H1}=\left(
\begin{array}{lccr}

-h+\frac{J}{4}\cos\Theta&-\frac{J}{4}\sin\Theta&\frac{J}{4}\sin\Theta&-\frac{J}{4}(1-\cos\Theta)\\

-\frac{J}{4}\sin\Theta&-\frac{J}{4}\cos\Theta&\frac{J}{4}(1+\cos\Theta)&-\frac{J}{4}\sin\Theta\\

\frac{J}{4}\sin\Theta&\frac{J}{4}(1+\cos\Theta)&-\frac{J}{4}\cos\Theta&\frac{J}{4}\sin\Theta\\

-\frac{J}{4}(1-\cos\Theta)&-\frac{J}{4}\sin\Theta&\frac{J}{4}\sin\Theta&h+\frac{J}{4}\cos\Theta
\end{array}
\right)
\end{equation}

After we do a rotation by $45$ degrees in the
$\{a\uparrow b\downarrow,a\downarrow b\uparrow\}$ subspace we
can apply perturbation theory (considering $J$ as a small
parameter), which yields (after subtracting $J$) the
eigenvalues : 

\begin{equation}
\label{EHeisenberg}
\left(
\begin{array}{llll}
h-\frac{J}{2}\sin^{2}\frac{\Theta}{2}&
0&
-J\cos^{2}\frac{\Theta}{2}&
-h-\frac{J}{2}\sin^{2}\frac{\Theta}{2}
\end{array}
\right)
\end{equation}

By matching the results in Eq. \ref{EHubbard1} to those in Eq. \ref{EHeisenberg} in
the $h\ll U$ limit (ignoring $O(\frac{t^2h}{U^2})$) we can
make the identification :
\begin{equation}
\label{J}
J=\frac{4t^2}{U}
\end{equation}
However, when the field is increased, the matching is not
perfect anymore, and the effective exchange parameter for the
ground state is reduced to 
\begin{equation}
J=\frac{4t^2}{U+h}
\label{newj}
\end{equation}
Thus the appearence of polarons decreases the effective
anti-ferromagnetic exchange between the carriers.

\section{Generalized Hubbard model}

\subsection{$2$-level Hubbard model with random fields}
\label{2levhub}
One can improve this analysis by considering a more realistic
model. The next simplest case is to consider two energy
levels $1$ and $2$ (energies $\epsilon_1$ and $\epsilon_2$)
on each atom, and to allow hopping $1\leftrightarrow
1$,$1\leftrightarrow 2$ and $2\leftrightarrow 2$ between
sites. Again we consider arbitrary fields $h_a$ and $h_b$.
The Hamiltonian is still given by Eq. \ref{fullHubb} with the summation for $i$ going over $1,2$. 
The number of states increases dramatically : we are dealing
now with a $28\times 28\ $ matrix ( $28=6\times (3+1)+4\times
1\ $since there are 6 pairs of different spatial states which
each can have spin $0$ or $1$, and 4 pairs of identical
states which can only have spin $0$). We need to concentrate
on the lowest energy states only, treating the rest
perturbatively. We ignore the $2\leftrightarrow 2$ hopping,
since it affects the lowest eigenvalues only to higher order
in PT. We are using again the simplifying assumption
$h_a=h_b=h$.

The lowest energy subspace can be identified as being spanned
by $\{a1\uparrow b1\uparrow, a1\uparrow b1\downarrow,
a1\downarrow b1\uparrow, a1\downarrow b1\downarrow\}$. By
applying second order degenerate PT in this subspace, we
obtain the following expressions for the eigenvalues : 
\begin{equation}
\label{EHubbard2}
\left(
\begin{array}{l}

2\epsilon_1+h-\frac{2t_{12}^2\cos^2\frac{\Theta}{2}}{\epsilon_2-\epsilon_1+U_{12}}-\frac{2t_{12}^2\sin^2\frac{\Theta}{2}}{\epsilon_2-\epsilon_1+U_{12}-h}-\frac{2t_{1a}^2\sin^2\frac{\Theta}{2}}{U_{11}-h}\\

2\epsilon_1-\frac{2t_{12}^2\cos^2\frac{\Theta}{2}}{\epsilon_2-\epsilon_1+U_{12}}-\frac{t_{12}^2\sin^2\frac{\Theta}{2}}{\epsilon_2-\epsilon_1+U_{12}-h}-\frac{t_{12}^2\sin^2\frac{\Theta}{2}}{\epsilon_2-\epsilon_1+U_{12}+h}\\

2\epsilon_1-\frac{4t_{11}^2\cos^2\frac{\Theta}{2}}{U_{1a}}-\frac{2t_{12}^2\cos^2\frac{\Theta}{2}}{\epsilon_2-\epsilon_1+U_{12}}-\frac{t_{12}^2\sin^2\frac{\Theta}{2}}{\epsilon_2-\epsilon_1+U_{12}-h}-\frac{t_{12}^2\sin^2\frac{\Theta}{2}}{\epsilon_2-\epsilon_1+U_{12}+h}\\

2\epsilon_1-h-\frac{2t_{12}^2\cos^2\frac{\Theta}{2}}{\epsilon_2-\epsilon_1+U_{12}}-\frac{2t_{12}^2\sin^2\frac{\Theta}{2}}{\epsilon_2-\epsilon_1+U_{12}+h}-\frac{2t_{11}^2\sin^2\frac{\Theta}{2}}{U_{11}+h}
\end{array}
\right)
\end{equation}

One can see that in the limit $t_{12}\rightarrow 0$ we obtain
the same results as in the $1$-level Hubbard model analyzed
in the beginning. This is a good consistency check. 

\subsection{Magnetic properties of the ground state}
\label{gsmagn}
By applying second order perturbation theory to the $2$-level
Hubbard model (and considering $h\gg t^2/U$), we therefore
obtain the following expression for the ground-state energy :
\begin{equation}
\label{HubbGS}
\begin{array}{c r r}
E_{GS}=2\epsilon_1-h-\frac{2t_{12}^2\cos^2\frac{\Theta}{2}}{\epsilon_2-\epsilon_1+U_{12}}-\\
\\
\frac{2t_{12}^2\sin^2\frac{\Theta}{2}}{\epsilon_2-\epsilon_1+U_{12}+h}-\frac{2t_{11}^2\sin^2\frac{\Theta}{2}}{U_{11}+h}
\end{array}
\end{equation}
The angle $\Theta$ between the two fields was regarded up to
this point as
an external parameter. All the calculations so far were done
under the
assumption that the magnetic field was fixed externally. We
must take
however into account the fact that the field is generated by
the actual
polaron, and that although the magnitude of the field is
fixed by the size
of the polaron, the direction is free to change. Therefore,
when
$T\rightarrow 0K$, $\Theta$ takes the value that minimizes
the energy.
Since $E_{GS}(\Theta)=const.+A(h)\sin^2(\frac{\Theta}{2})$,
with
$$A(h)=\frac{2t_{12}^2}{\epsilon_2-\epsilon_1+U_{12}}-\frac{2t_{12}^2}{\epsilon_2-\epsilon_1+U_{12}+h}-\frac{2t_{11}^2}{U_{11}+h}$$
the two values that minimize the value of the energy are
$\Theta=0$ and
$\Theta=\pi$, depending of the sign of the factor $A(h)$. We
can regard
$A$, which represents the energy difference between the
ferromagnetic and
the anti-ferromagnetic configurations, as an effective
exchange constant.
For small
values of the polaron field it is the anti-ferromagnetic
state that dominates, whereas if we increase the polaron
field the ground state of the system becomes ferromagnetic. 

In order to get an idea of what parameters are essential for
the transition, let us solve $A(h)=0$, which is just a
quadratic equation. The critical field is given by the only
acceptable (positive) root: 
\begin{equation}
h_{c}=U_{11}\frac{-1+\alpha\beta+\sqrt{1-2\alpha\beta+\alpha^2\beta^2+4\alpha\beta^2}}{2}
\end{equation}
where we have defined $\alpha=(\frac{t_{11}}{t_{12}})^2$ and
$\beta=\frac{U_{12}+\epsilon_2-\epsilon_1}{U_{11}}$. In the
one-level limit, thus, the transition disappears (we have
$\alpha\rightarrow\infty\Rightarrow h\rightarrow\infty$). The
ferromagnetic configuration happens then as a consequence of
the local magnetic fields \emph{and} of the hopping to
excited states. As a side remark we observe that while
$\beta$ depends entirely on the type of dopant used, $\alpha$
depends both on the type of dopant \emph{and} on the dopant
density (since all the transition probabilities depend on the
separation of the atoms). Thus we may have a transition from
a polaron ferromagnetic to an anti-ferromagnetic ground state
configuration as the density is varied. It turns out that for
the $2$-level (hydrogenic ground state and any excited state)
approximation, $\alpha$ can take virtually any value from
$\alpha=0$ for $0$ separation, to $\alpha \rightarrow \infty$
when the separation becomes infinite.

\subsection{Many-level Hubbard Model}
\label{mlevhub}
One can generalize the results presented above about the
ground-state of a $2$-level Hubbard model to the case where
any number of bound excited states exists. Consider
$|\psi_i\uparrow\rangle$ and $|\psi_i\downarrow\rangle$ (with
energies $\epsilon_i$) to be the $1$-electron states, indexed
by the subscript $i$ ($i=0$ for the ground-state). We
consider the ground-state to be non-degenerate, however
allowing arbitrary degeneracies for all the other states. The
ground-state of the two-electron system in a magnetic field
is then : $a\psi_0\downarrow b\psi_0\downarrow$, with energy
(in the $0$-th order) $2\epsilon_0-h$. We allow the hopping
of an electron from the ground state of one atom to any state
of the other atom, the coupling constants between $a\psi_0$
and $b\psi_i$ being given by $t_{0i}$. Expression \ref{fullHubb} gives the right Hamiltonian with the summation for $\alpha$ being over $a,b$, corresponding to only two interacting polarons. 

The couplings of the ground-state  $a\psi_0\downarrow
b\psi_0\downarrow$ to various other states are given in Table
\ref{coupltable}. Applying second order non-degenerate PT, we
obtain for the ground state energy : 
\begin{equation}
\label{HubbGSmany}
\begin{array}{c r r}
E_{GS}=2\epsilon_0-h-\frac{2t_{00}^2\sin^2\frac{\Theta}{2}}{U_{00}+h}-(\sum_i{\frac{2t_{0i}^2}{\epsilon_i-\epsilon_0+U_{0i}}})\cos^2\frac{\Theta}{2}-(\sum_i{\frac{2t_{0i}^2}{\epsilon_i-\epsilon_0+U_{0i}+h}})\sin^2\frac{\Theta}{2}
\end{array}
\end{equation}

We can apply the same kind of analysis as for  the two-level
case above, however $A(h)$, the effective exchange constant
whose sign dictates the magnetic configuration now becomes:
\begin{equation}
\label{ah}
A(h)=
\sum_i{\frac{2t_{0i}^2}{\epsilon_i-\epsilon_0+U_{0i}}}-\sum_i{\frac{2t_{0i}^2}{\epsilon_i-\epsilon_0+U_{0i}+h}}-\frac{2t_{00}^2}{U_{00}+h}
\end{equation}
\begin{table}
\begin{tabular}{|c|c|c|}\hline
\emph{State} & \emph{Coupling} & \emph{Energy diff.} \\
\hline
$a(b)\psi_0\downarrow a(b)\psi_0\uparrow$ & 
$\pm i t_{00} \sin(\frac{\Theta}{2})$ & 
$U_{00}+h$ \\ \hline
$a(b)\psi_i\uparrow a(b)\psi_0\downarrow,\
a(b)\psi_i\downarrow a(b)\psi_0\uparrow$ & $\pm i t_{0i}
\sin(\frac{\Theta}{2})$ & $\epsilon_i-\epsilon_0+U_{0i}+h$ \\
\hline
$a(b)\psi_i\downarrow a(b)\psi_0\downarrow,\
a(b)\psi_i\uparrow a(b)\psi_0\uparrow$ & $\pm t_{0i}
\cos(\frac{\Theta}{2})$ & $\epsilon_i-\epsilon_0+U_{0i}$ \\
\hline
\end{tabular}
\caption{The couplings of the ground state of the many-level
Hubbard model to various excited states} \label{coupltable}
\end{table}
The value of the critical field above which the ferromagnetic
configuration becomes energetically favorable is again given
by the
equation $A(h_c)=0$, however this cannot be solved
analytically anymore.
\begin{figure}
\epsfbox{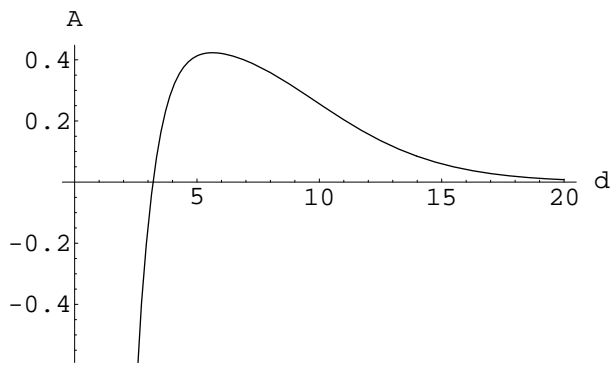}
\caption{The effective exchange $A$ as a function of the
dopant separation
$d$. The polaron field was taken as $h=0.3$Ry } 
\label{ah1}
\end{figure}
\begin{figure}
\epsfbox{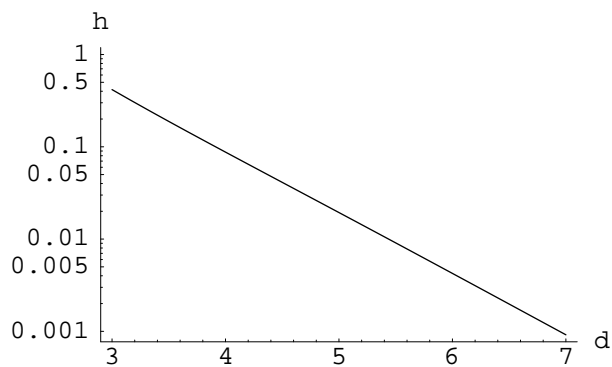}
\caption{The critical field $h_c$ (in Ry) \emph{vs.} the
dopant separation
$d$ (in $a_B$). The Log-Linear plot reveals the exponential
behaviour of the critical field on the dopant separation.}
\label{hcrit}  
\end{figure}
\subsection{Application to Hydrogenic Centers}
\label{hydcen}
We can understand the details of this change of magnetic
configuration better if we derive the actual couplings and
Coulomb terms from a simplified model of the dopant. As
discussed in the introduction, one can use simple Hydrogenic
models : two H atoms separated by a distance $\rho$ in
magnetic fields $\mathbf{h_a,\ h_b}$. We obtained the
2-center overlap integrals ($t_{0i}$) by using explicit
machine-readable formulas that have been constructed by
applying symbolic calculation to the $\zeta$-function method
of Barnett and Coulson. The mathematical formalism has been
described originally here \cite{barnett:1}. The symbolic
calculations are described here
\cite{barnett:2}\cite{barnett:3} and the work cited therein.
The evaluation of $1$-center Coulomb integrals ($U_{0i}$) was
done following an approach described in some textbooks\cite{slater:qtms}.

We considered Hydrogenic centers, and included
transitions and Coulomb interaction energies between
the ground state $1s$ and the states $1s$, $2s$, $2p_z$,
$3s$, $3p_z$, and $3d$. 
Figure \ref{ah1} plots the effective exchange $A$
as a function of separation $d$ between the two centers,
for a fixed value of the polaron effective field,
$h=0.3$Ry, which is still small
compared to the Rydberg. It can be seen that 
the effective exchange $A$ becomes positive at a
certain dopant
separation $d$ and thus the favorable configuration becomes
ferromagnetic.
Figure \ref{hcrit} plots the critical field $h_c$ as a
function of the
dopant separation $d$. For $d>4a_B$, which is true for
typical
experimental doping densities, the minimal value of the
polaron field
that will provide a ferromagnetic interaction becomes
reasonable (a few
tenth of a Rydberg) and therefore we can conclude that our
model predicts
ferromagnetic interractions between polarons in DMS.

\section{Spin Hamiltonian for Moderate Fields}
\label{modif}
Coming back to our fixed magnetic field model, we note that
both the $1$- and $2$-level Hubbard models agree with the
Hamiltonian of Eq. {\ref{heis}}, containing the standard
Heisenberg exchange and Zeeman terms when contributions of order
$O(\frac{t^2h}{U^2})$ are ignored. However when those contributions
are taken into account, the Eq.
\ref{heis} does not provide the right solutions anymore. The
question to be asked is whether it is possible to modify the
Hamiltonian so as to have agreement up to
$O(\frac{t^2h}{U^2})$. It turns out that this is indeed
possible. 

If we expand the terms in Eq. \ref{EHubbard1} we obtain the
energies :
\begin{equation}
\label{newEHubbard1}
\left(
\begin{array}{llll}

2\epsilon+h-\frac{2t^{2}\sin^{2}\frac{\Theta}{2}}{U}-\frac{2t^{2}h\sin^{2}\frac{\Theta}{2}}{U^2}&
2\epsilon&
2\epsilon -\frac{4t^{2}\cos^{2}\frac{\Theta}{2}}{U}&

2\epsilon-h-\frac{2t^{2}\sin^{2}\frac{\Theta}{2}}{U}+\frac{2t^{2}h\sin^{2}\frac{\Theta}{2}}{U^2}
\end{array}
\right)
\end{equation}

We need to add some small correction to Eq. \ref{heis} that
is linear in the fields and reproduces the above structure.
There are several ways of doing this, the simplest being to
add a term of the form $\mathbf{h_a\cdot
s_b}+\mathbf{h_b\cdot s_a}$ or $\mathbf{h\cdot s_a\times
s_b}$ or $\mathbf{h_a\cdot s_a}+\mathbf{h_b\cdot s_b}$ (or
any linear combination thereof). It turns out that the
Hamiltonian 
\begin{equation}
\label{newheis}
\begin{array}{lc}
H_{H2}=(1-\alpha)(\mathbf{h_a\cdot s_a}+\mathbf{h_b\cdot
s_b})+J\mathbf{s_a\cdot s_b}+\alpha(\mathbf{h_a\cdot
s_b}+\mathbf{h_b\cdot s_a})
\end{array}
\end{equation}
reproduces the right structure. In the basis $\{a\uparrow
b\uparrow,a\uparrow b\downarrow,a\downarrow
b\uparrow,a\downarrow b\downarrow\}$ it becomes :

\[
\label{newHHeisenberg}
\left(
\begin{array}{lccr}
-(1-\alpha)h+(\frac{J}{4}-\alpha
h)\cos\Theta&-(\frac{J}{4}-\frac{\alpha
h}{2})\sin\Theta&(\frac{J}{4}-\frac{\alpha
h}{2})\sin\Theta&-\frac{J}{4}(1-\cos\Theta)\\
-(\frac{J}{4}-\frac{\alpha h}{2})\sin\Theta&-\frac{J}{4}\cos\Theta&\frac{J}{4}(1+\cos\Theta)&-(\frac{J}{4}+\frac{\alpha
h}{2})\sin\Theta\\
(\frac{J}{4}-\frac{\alpha
h}{2})\sin\Theta&\frac{J}{4}(1+\cos\Theta)&-\frac{J}{4}\cos\Theta&(\frac{J}{4}+\frac{\alpha
h}{2})\sin\Theta\\
-\frac{J}{4}(1-\cos\Theta)&-(\frac{J}{4}+\frac{\alpha
h}{2})\sin\Theta&(\frac{J}{4}+\frac{\alpha
h}{2})\sin\Theta&(1-\alpha)h+(\frac{J}{4}+\alpha h)\cos\Theta
\end{array}
\right)
\]

After doing the necessary $45$ degree rotation in the
$\{a\uparrow b\downarrow,a\downarrow b\uparrow\}$ subspace
the matrix becomes suitable for PT and it yields the
eigenvalues (to first order in $J$ and $\alpha$ and after
subtracting a common $J$):
\begin{equation}
\label{newEHeisenberg}
\left(
\begin{array}{llll}
h-\frac{J}{2}\sin^{2}\frac{\Theta}{2}-2\alpha
h\sin^2\frac{\theta}{2}&
0&
-J\cos^{2}\frac{\Theta}{2}&
-h-\frac{J}{2}\sin^{2}\frac{\Theta}{2}+2\alpha
h\sin^2\frac{\theta}{2}
\end{array}
\right)
\end{equation}

By matching the results in Eq. \ref{newEHeisenberg} to Eq.
\ref{newEHubbard1} we obtain for the parameters of the
modified Hamiltonian : 
\begin{equation}
\label{parameters1}
\begin{array}{ll}
J=\frac{4t^2}{U}&
\alpha=\frac{t^2}{U^2}
\end{array}
\end{equation} 
One can also expand the result for the $2$-level Hubbard
model to get a better estimate for the parameters. In that
case one obtains : 
\begin{equation}
\label{parameters2}
\begin{array}{ll}

J=\frac{4t_{11}^2}{U_{11}}&\alpha=\frac{t_{11}^2}{U_{11}^2}+\frac{t_{12}^2}{(\epsilon_2-\epsilon_1+U_{12})^2}
\end{array}
\end{equation}  

\section{Conclusion}

The calculations presented above lead us to two conclusions : 
\newcounter{conc}
\begin{list}
{\Roman{conc}}{\usecounter{conc}}
\item They confirm once more the fact
that the polarons formed in Dilute Magnetic Semiconductors
can interact ferromagnetically for certain dopant densities
and types. This extends the results obtained by
\cite{bhatt:polarons}$^,$\cite{bhatt1:polarons} in the limit that
the polarons have an important overlap to the situation where
the two polarons do not overlap at all. Thus, the two
qualitatively distinct effects combine in order to generate
an effective ferromagnetic interaction of the bound magnetic
polarons in a DMS. 
\item For the case where two $1$-electron atoms are placed in fixed,
but nonparallel external magnetic fields, 
the standard model with Heisenberg exchange and Zeeman terms
is not a suitable approximation.Instead, the calculations 
above give
a correction, with which we are able to reproduce the
correct spectrum in the moderately high field domain $(t\ll
h\ll U)$. The correction represents an effective mixing of
the fields at the two sites, which can be intuitively
understood as a ``transfer'' of the field from one site to
the other by the hopping electron. In the high field domain
$(h<U)$, this correction is not valid anymore and the correct
model is a Heisenberg Hamiltonian with a field-dependent
exchange constant.
\end{list}

We will return now to Conclusion I with a few remarks. The change from antiferromagnetic to ferromagnetic
effective coupling in the presence of strong local fields
in the generalized Hubbard model naturally leads one to
the question whether this would actually occur in an
exact calculation. We believe it does, though the
parameter values for the change are likely to be different.
To explain our ``belief'', we consider the case of the
hydrogen molecule problem in zero field, where the
issue of the effective Heisenberg Hamiltonian has been
thoroughly discussed \cite{Heisenberg:}$^,$  \cite{Heitler:} $^,$ \cite{Herring:}$^,$ \cite{Herring1:}.

\begin{figure}
\epsfbox{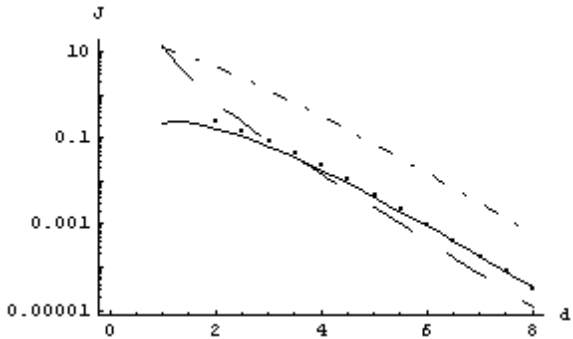}
\caption{The effective exchange parameter as a function of reduced distance $(r/a_B)$ for Herring-Flicker (solid), Kolos-Wolniewicz (dots), Heitler-London (dashed) and Hubbard model (dot-dashed).} 
\label{Js}
\end{figure}

In Fig. \ref{Js} we plot the effective exchange parameter as
a function of $d=(r/a_B)$ in the range 1-8, as calculated in
four different ways. The solid line represents
the Herring-Flicker\cite{Herring:} (HF)  result,
$ J_{HF} (r) = 1.636 (r/a_B)^{5/2}\exp(-2r/a_B) $ Ry,
which is asymptotically exact (in the sense that
$ J_{HF} (r)/J_{exact} (r) \rightarrow 1 $ as
$ r/a_B \rightarrow \infty $ ), 
while the dots are the numerically converged results
of Kolos and Wolniewicz \cite{kolos:woln} (KW).
Both show that $J(r)$ is positive (antiferromagnetic)
for all $r$. The dashed line is the result of the
Heitler-London \cite{Heitler:} (HL) approximation,
which, though clearly not exact, works reasonably well
on this logarithmic plot for the range shown. It should
be noted, however, that while the HL result has the
right sign of $J(r)$ for the range of $r/a_B$ shown,
it incorrectly predicts a negative $J(r)$ at large $r/a_B$
because it does not take into account polarization corrections
to the ground state hydrogenic wavefunction. Finally,
as the dot-dashed curve, is the standard Hubbard model result,
$ J(r) = 4 t^2(r)/U $, with $ t(r) = 2 (1 + r/a_B) \exp (-r/a_B)$
Ry and $U= 5/4$ Ry , calculated within the ground state
approximation for the hydrogen wavefunctions
(the generalized Hubbard model would give the same result
in this case without external fields, as adding excited states
does not alter the second order splitting between the lowest
singlet and triplet states).

As can be clearly seen, the Hubbard approximation overestimates
$ J(r) $ by a large factor (this qualitative fact does not
change with more refined estimates for $t(r)$ and $U$ ).
The reason for the larger exchange is that Hubbard, and
Hubbard-like approximations, consider only the kinetic exchange
(due to the hopping process), which is antiferromagnetic,
and neglect coulomb exchange which tends to favor ferromagnetism.
(Such a split is often used in literature\cite{Fazekas:}).
The latter {\em is} included in the ``exact" HF/KW treatment,
as well as the HL approximation, resulting in a much lower value
net (antiferromagnetic) exchange.

We expect that inclusion of local magnetic fields 
$ h_i $ of equal magnitude ($h$), which couple
{\em only} to the electron spin 
is properly captured on a qualitative/semi-quantitative
level by the extended Hubbard model.
Therefore inclusion of such fields in a more accurate
model will also
result in a movement of the kinetic exchange towards
ferromagnetism; consequently, the overall exchange will
change over to ferromagnetic at some value of $h$.
If we just add a field independent 
(ferromagnetic) coulomb term to the
kinetic exchange of the generalized Hubbard model,
the change from antiferromagnetism to ferromagnetism
would be expected at lower values of $h$ than we have
calculated, and make the effect we consider more relevant
for DMS systems.

We caution, though, that for the purely hydrogenic problem
with local fields, one must take into account
the effect of the magnetic field on the orbital wavefunction
as well, and that will certainly affect the results, at
least quantitatively. In the case of DMS, the local fields
represent exchange fields due to interaction of hydrogenic
states with local atomic states of the magnetic ion (Mn),
and therefore their orbital effect is {\em not} the same as that
of external magnetic fields in the $H_2$ problem. Nevertheless,
we expect these to have some effect on the orbital part
of the wavefunction of the hydrogenic impurity \cite{Mohan:}, which would have at least
a quantitative effect on our results.

\section{Acknowledgments}
This research was supported by NSF DMR-9809483.

\end{document}